\pdfoutput=1
\documentclass[]{smule}
\usepackage[round,authoryear]{natbib}
\usepackage{hyperref}
\usepackage[colorinlistoftodos]{todonotes}
\usepackage{colortbl}
\usepackage{nicematrix}
\usepackage{amssymb}
\usepackage{amsmath}
\usepackage{booktabs}
\usepackage{adjustbox}
\usepackage{soul}
\usepackage[inline]{enumitem}
\usepackage{booktabs}
\usepackage{color}
\usepackage{xcolor}
\usepackage{bbding}
\usepackage{listings}
\usepackage{multicol}
\usepackage{xspace}
\usepackage{tablefootnote}
\usepackage{libertinus}

\makeatletter
\AtBeginDocument{

}
\makeatother

\title{FlowSynth: Instrument Generation Through Distributional Flow Matching and Test-Time Search}

\author[2\dagger]{Qihui Yang}
\author[1]{Randal Leistikow}
\author[1]{Yongyi Zang}

\affiliation[1]{Smule Labs}
\affiliation[2]{University of California, San Diego}

\contribution[\dagger]{Work done during internship at Smule}

\abstract{
Virtual instrument generation requires maintaining consistent timbre across different pitches and velocities, a challenge that existing note-level models struggle to address. We present \textbf{FlowSynth}, which combines distributional flow matching (DFM) with test-time optimization for high-quality instrument synthesis. Unlike standard flow matching that learns deterministic mappings, DFM parameterizes the velocity field as a Gaussian distribution and optimizes via negative log-likelihood, enabling the model to express uncertainty in its predictions. This probabilistic formulation allows principled test-time search: we sample multiple trajectories weighted by model confidence and select outputs that maximize timbre consistency. FlowSynth outperforms the current state-of-the-art TokenSynth baseline in both single-note quality and cross-note consistency. Our approach demonstrates that modeling predictive uncertainty in flow matching, combined with music-specific consistency objectives, provides an effective path to professional-quality virtual instruments suitable for real-time performance.
}

\begin{document}

\maketitle

\section{Introduction}
\label{sec:intro}

Modern music production demands precise control over timbre, pitch, and dynamics at the note level. While recent advances in flow matching~\citep{lipman2023flowmatchinggenerativemodeling} have enabled high-quality music generation through models \citep{musicflow, ning2025diffrhythmblazinglyfastembarrassingly}, maintaining consistent timbre across an instrument's range remains a fundamental challenge for note-level generation. This gap limits the adoption of AI-generated instruments in professional music production, where musicians expect reliable timbral behavior comparable to physical instruments or high-quality sample libraries.

Virtual instrument generation, the task of synthesizing individual notes with specified pitch, velocity, and timbre for real-time playability, has evolved significantly from early neural approaches. NSynth \citep{engel2017neuralaudiosynthesismusical} pioneered learned instrument synthesis, while DDSP \citep{engel2020ddspdifferentiabledigitalsignal} introduced differentiable signal processing for more controllable generation. The current state-of-the-art, TokenSynth \citep{kim2025tokensynthtokenbasedneuralsynthesizer}, employs transformer architectures with CLAP embeddings \citep{wu2024largescalecontrastivelanguageaudiopretraining} to achieve reasonable timbre consistency across 88 piano keys. However, even these advanced models exhibit timbral drift: a "warm piano" with a fundamental frequency of C3 (131 Hz) may sound metallic at C6 (1046 Hz), despite identical text conditioning. This inconsistency partly stems from the deterministic nature of current generation pipelines, which lack mechanisms to explore multiple solutions and select for consistency, a limitation noted in studies on timbre disentanglement \citep{luo2019learningdisentangledrepresentationstimbre}. Meanwhile, the broader generative modeling community has explored probabilistic approaches to address similar challenges. Variational Rectified Flow Matching \citep{guo2025variationalrectifiedflowmatching} introduces latent variables for capturing multi-modal velocity fields in graph generation, while energy-weighted flow matching \citep{woo2024iteratedenergybasedflowmatching} uses importance weighting for challenging distributions. In parallel, test-time compute scaling has demonstrated remarkable gains in language models \citep{snell2024scalingllmtesttimecompute}, with best-of-N sampling and its variants proving effective across domains \citep{mudgal2024controlleddecodinglanguagemodels, gui2024bonbonalignmentlargelanguage}. However, these advances have not been adapted to address the specific challenge of cross-note timbre consistency in virtual instruments.

We introduce FlowSynth, which addresses timbre consistency through distributional flow matching (DFM), a novel formulation that learns distributions over velocity fields rather than point estimates. While Variational Rectified Flow Matching \citep{guo2025variationalrectifiedflowmatching} also models velocity field uncertainty through latent variables optimized with an evidence lower bound (ELBO) objective assuming unit Gaussian priors, DFM takes a more direct approach: we parameterize the velocity field itself as a Gaussian distribution with learned mean and variance, optimized via negative log-likelihood (NLL). NLL optimization without unit Gaussian prior assumptions enables our model to learn problem-specific uncertainty patterns. The NLL objective naturally encourages higher-variance predictions in regions of genuine ambiguity (where multiple valid velocities exist), while maintaining confident predictions elsewhere. This learned uncertainty map proves crucial for test-time search, allowing us to explore more extensively in ambiguous regions while preserving high-confidence predictions, ultimately enabling the principled trajectory sampling that drives our consistency improvements.

At test time, we leverage this learned uncertainty through a temperature-controlled sampling strategy that scales with computational budget. Drawing inspiration from extreme value theory, we design a sublinear temperature schedule that balances exploration with stability. For each note generation after the first, we sample multiple trajectories weighted by the model's confidence and evaluate them against a timbre consistency objective—transforming generation from a single-shot process to an optimization problem where additional compute directly improves instrument quality.

FlowSynth builds on established audio generation architectures, employing a Diffusion Transformer (DiT) \citep{peebles2023scalablediffusionmodelstransformers} with adaptive layer normalization for conditioning, similar to recent models like AudioX \citep{tian2025audioxdiffusiontransformeranythingtoaudio}. We leverage CLAP encoders \citep{wu2024largescalecontrastivelanguageaudiopretraining} for text-audio alignment and adopt the VAE from DiffRhythm \citep{ning2025diffrhythmblazinglyfastembarrassingly} for efficient latent space modeling. This architectural foundation, combined with our distributional formulation, delivers substantial improvements: even without test-time search, FlowSynth surpasses TokenSynth in audio quality and timbre consistency, with comparable prompt adherence when using unconditional search. The true strength of DFM emerges with test-time compute scaling: increasing the search budget by $8\times$ improves timbre consistency by $13$\% and prompt adherence by $246$\%, demonstrating that our uncertainty-aware approach effectively converts compute into quality gains.

Our contributions are: (1) Distributional flow matching (DFM), a formulation using negative log-likelihood optimization that captures predictive uncertainty for improved generation; (2) A test-time optimization framework combining confidence-weighted sampling with music-specific consistency objectives; (3) FlowSynth, achieving state-of-the-art results in virtual instrument generation with demonstrated compute scaling benefits. By combining uncertainty-aware generation with domain-specific optimization, we provide a practical path toward professional-quality virtual instruments that maintain consistent timbre across their entire range.

\begin{figure*}[h]
    \centering
    \includegraphics[width=\linewidth]{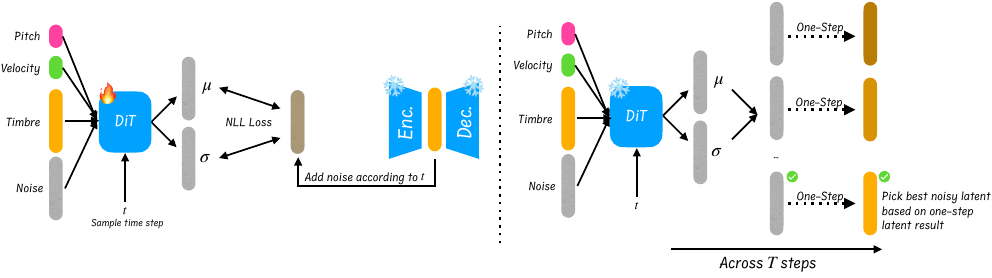}
    \caption{Proposed distributional flow matching (DFM) architecture. \textbf{(Left) Training phase:} The diffusion transformer (DiT) learns to predict the instantaneous velocity field by projecting latent representations into mean ($\mu$) and variance ($\sigma$) parameters of a Gaussian distribution. Negative log-likelihood (NLL) loss is minimized between the ground-truth latent at current timestep and this learned distribution. \textbf{(Right) Inference phase:} Multiple velocity field samples are drawn from the learned Gaussian distribution at each generation timestep $t$. These samples undergo iterative one-step generation until the final timestep $T$, where a search metric evaluates all candidate latents to select the optimal output for the current timestep.
    }
    \label{fig:teaser}
\end{figure*}

\section{Distributional Flow Matching}
\label{sec:method}

Standard flow matching learns deterministic mappings from source to target distributions, providing no mechanism to express predictive uncertainty or explore alternative generation paths. We propose Distributional Flow Matching (DFM), which reformulates velocity field prediction as learning a probability distribution over possible velocities, enabling uncertainty-aware generation and principled test-time exploration.

\subsection{Probabilistic Velocity Field Formulation}

In conventional flow matching, a neural network $f_\theta$ predicts a deterministic velocity field $v_\theta(x_t, t)$ that transports samples along an optimal transport path. We extend this framework by modeling the velocity field as a conditional probability distribution:
\begin{equation}
    p(v | x_t, t, c) = \mathcal{N}(v; \mu_\theta(x_t, t, c), \sigma^2_\theta(x_t, t, c)I),
\end{equation}
where $x_t$ denotes the state at time $t$, $c$ represents optional conditioning information, and $I$ is the identity matrix. The neural network jointly predicts the distribution parameters:
\begin{equation}
    [\mu_\theta(x_t, t, c), \log \sigma^2_\theta(x_t, t, c)] = f_\theta(x_t, t, c).
\end{equation}
We parameterize the log-variance to ensure numerical stability and strictly positive variance values through the exponential transformation.

\subsection{Training via Negative Log-Likelihood}

Unlike variational approaches that introduce auxiliary latent variables and optimize an evidence lower bound, DFM directly optimizes the negative log-likelihood of ground-truth velocities under the predicted distribution. Given training pairs $(x_t, v_t)$ sampled from the optimal transport path between data distributions, we minimize:
\begin{equation}
    \mathcal{L}_{\text{DFM}} = \mathbb{E}_{(x_t, t, v_t)} \left[ \frac{d}{2} \log \sigma^2_\theta + \frac{\| v_t - \mu_\theta \|^2}{2 \sigma^2_\theta} \right],
\label{eq:dfm-loss}
\end{equation}
where $d$ is the velocity dimensionality. This objective exhibits two properties: (1) the variance acts as an adaptive weighting mechanism, where the model can express uncertainty through higher variance predictions, which down-weights the reconstruction penalty in ambiguous regions; (2) the log-variance penalty prevents degenerate solutions, encouraging confident predictions where the transport path is unambiguous. The learned variance structure allows for higher variance to emerge at distribution boundaries and multi-modal regions where multiple valid velocities exist, while maintaining low variance along clear transport paths. 

\subsection{Uncertainty-Guided Sampling}

During inference, we leverage the learned uncertainty to guide trajectory sampling. Rather than deterministic integration or arbitrary noise injection, we sample velocities proportional to the model's confidence:
\begin{equation}
v_{\text{sampled}} = \mu_\theta(x_t, t, c) + \tau \cdot \sigma_\theta(x_t, t, c) \cdot \boldsymbol{\epsilon},
\label{eq:uncertainty-sampling}
\end{equation}
where $\boldsymbol{\epsilon} \sim \mathcal{N}(0, I)$ and $\tau$ is a temperature parameter controlling exploration strength. This formulation ensures exploration is concentrated in regions of genuine ambiguity while preserving high-fidelity generation where the model is confident.

The temperature parameter enables a spectrum of behaviors: $\tau = 0$ recovers deterministic generation using the mean prediction, small values ($\tau \approx 0.1$) enable local refinement, and moderate values ($\tau \approx 1$) allow broader exploration while respecting the learned uncertainty structure. Crucially, because the variance is learned rather than prescribed, the effective exploration adapts to the inherent difficulty of each generation step.

\section{FlowSynth}
\label{sec:flowsynth}

We instantiate DFM for virtual instrument generation through FlowSynth, a system that combines uncertainty-aware generation with domain-specific consistency objectives to produce high-quality, playable instruments.

\subsection{Architecture}

FlowSynth employs a Diffusion Transformer (DiT) \citep{peebles2023scalablediffusionmodelstransformers} backbone operating in a learned latent space. We adopt the variational autoencoder from DiffRhythm \citep{ning2025diffrhythmblazinglyfastembarrassingly}, which provides an efficient latent representation. The transformer processes flattened latent sequences through 24 layers with hidden dimension 1024, using rotary position embeddings for spatial awareness.

The network outputs both velocity mean and log-variance through separate projection heads from the final transformer layer. To prevent variance collapse during early training, we initialize the log-variance head to output $\log(0.1)$ and apply gradient clipping specifically to variance predictions.

\subsection{Conditioning Signals}
Virtual instrument generation requires precise control over multiple attributes. FlowSynth integrates three conditioning signals:

\textbf{Text Description.} We encode text prompts using a frozen CLAP encoder \citep{wu2024largescalecontrastivelanguageaudiopretraining}, projecting the 512-dimensional embeddings to the model's hidden dimension. This provides semantic control over timbre characteristics (e.g., "warm vintage piano").

\textbf{Pitch Control.} MIDI pitch values (21-108 for piano range) are embedded through a learned lookup table, enabling precise frequency control while maintaining timbre consistency across octaves.

\textbf{Velocity Dynamics.} MIDI velocity (1-127) is similarly embedded, controlling amplitude and timbral brightness as in acoustic instruments, where harder strikes produce both louder and brighter tones.

These signals combine through adaptive layer normalization (AdaLN), where the summed conditioning vector modulates each transformer block via learned scale ($\gamma$), shift ($\beta$), and gating ($\alpha$) parameters:
\begin{equation}
    x' = x + \alpha \odot \text{Block}(\text{LN}(x) \cdot (1 + \gamma) + \beta).
\end{equation}

\subsection{Test-Time Optimization for Timbre Consistency}
We design a temperature schedule that scales with computational budget while preventing instability:
\begin{equation}
\tau(N) = \min\left(\tau_{\max}, \tau_0 \sqrt{2 \ln(N+1)}\right),
\label{eq:temp-flowsynth}
\end{equation}
where $N$ is the number of candidate trajectories. The $\sqrt{2\ln(N+1)}$ scaling follows from extreme value theory, ensuring the expected maximum deviation grows sublinearly with samples. We find $\tau_0 = 0.01$ and $\tau_{\max} = 0.08$ work robustly across instruments.

For each generated note after the first, we sample $N$ trajectories and evaluate timbre consistency using CLAP audio embeddings:
\begin{equation}
s_i = \frac{1}{M-1} \sum_{j \neq i} \text{cos}(\phi_{\text{CLAP}}(x_i), \phi_{\text{CLAP}}(x_j)),
\end{equation}
where $M$ is the number of previously generated notes in the instrument. This encourages selection of trajectories that maintain consistent timbral characteristics across the pitch range. We balance consistency with text prompt adherence by computing:
\begin{equation}
s_{\text{total}} = \lambda \cdot s_{\text{consistency}} + (1-\lambda) \cdot s_{\text{prompt}},
\end{equation}
where $s_{\text{prompt}}$ measures alignment between generated audio and text embedding. We set $\lambda = 0.7$ to prioritize consistency while maintaining semantic control.

\section{Experiments}

\subsection{Dataset}

We train and evaluate on the NSynth dataset \citep{engel2017neuralaudiosynthesismusical}, which contains 305,979 musical notes from 1,006 instruments across 11 instrument families. Each 4-second audio sample is annotated with pitch (MIDI notes 21-108), velocity (5 levels: 25, 31, 50, 75, 100), and instrument family. We use the standard train/validation/test split with 289,205/12,678/4,096 samples respectively. For multi-note evaluation, we construct virtual instruments by sampling multiple pitches from the same source instrument, ensuring consistent timbre ground truth. We focus on the 88-key piano range (A0 to C8) as it represents the most demanding use case for timbral consistency.

\subsection{Evaluation Metrics}
We employ complementary metrics to assess generation quality, controllability, and consistency:

\textbf{Audio Quality.} Fréchet Audio Distance (FAD) \citep{kilgour2019frechetaudiodistancemetric} measures distributional similarity between generated and real audio using VGGish embeddings. Lower values indicate better quality.

\textbf{Pitch Accuracy.} Mean Absolute Deviation (MAD$_{\text{pitch}}$) quantifies pitch control by comparing the fundamental frequency of generated audio (extracted via YIN algorithm\citep{deCheveign2002YINAF} against the target MIDI pitch. We report deviation in cents (1/100th semitone).

\textbf{Prompt Adherence.} CLAP score \citep{wu2024largescalecontrastivelanguageaudiopretraining} measures semantic alignment between text prompts and generated audio using cosine similarity in the CLAP embedding space. Higher values indicate better text-to-audio correspondence.

\textbf{Timbre Consistency Loss.} Following InstrumentGen\citep{nercessian2023instrumentgengeneratingsamplebasedmusical}, we compute the timbre consistency loss using a two-level averaging strategy to ensure scale-invariant evaluation across different numbers of audio clips and varying temporal lengths:

The timbre consistency loss is calculated using a two-level averaging process to ensure fair comparisons across different conditions. First, for each pair of modified MFCC representations ($y_i$, $y_j$) in a group, we compute a pairwise distance $d_{ij} = (1/D) \|y_i - y_j\|_1$, which is normalized by the feature dimensionality $D$ to remove dependence on feature vector length. Second, we average these distances across all $\frac{K(K-1)}{2}$ unique pairs in the group of $K$ clips to obtain the final loss, $L_{\text{timbre}} = (2/[K(K-1)]) \sum_{i<j} d_{ij}$. This group-level normalization makes the loss magnitude independent of the number of clips in the group, ensuring a consistent metric.

\textbf{Perceptual Quality.} Multi-Scale Spectral (MSS) loss \citep{10319088} evaluates spectral fidelity at multiple time scales. For multi-note evaluation, we report the onset F-score to assess attack transient quality, crucial for playability.

\subsection{Experimental Setup}

\textbf{Baselines.} We compare against TokenSynth \citep{kim2025tokensynthtokenbasedneuralsynthesizer}, the current state-of-the-art for controllable instrument generation. We evaluate our model in multiple configurations: (1) deterministic generation without search ($\tau=0$), (2) unconditional search with varying budgets (N=\{8, 16, 32\}), and (3) guided search using task-specific objectives.

\textbf{Training Configuration.} Training uses the AdamW optimizer with learning rate $10^{-4}$ and cosine schedule over 500K steps. We train on 8 NVIDIA A100 GPUs with batch size 32 per device. The flow matching process uses linear interpolation between Gaussian noise and data with 1000 discrete timesteps during training. Variance predictions are regularized with gradient clipping at 1.0 to prevent instability.

\textbf{Inference Configuration.} We employ a dopri5 solver with 16 integration steps for generation. For test-time search, we sample trajectories using the temperature schedule in Eq.~\ref{eq:temp-flowsynth}. The search evaluates candidates in parallel batches, with early stopping when the consistency metric plateaus (typically after 10-20 candidates).

\begin{table}[ht]
\centering
\caption{Single Note Results without guided search}
\begin{tabular}{l c c c c c}
\toprule
\textbf{Model} & \textbf{MAD}$_{pitch}$$\downarrow$& \textbf{MSS}$\downarrow$& \textbf{CLAP}$\uparrow$& \textbf{FAD}$_{vgg}$$\downarrow$& \textbf{TCC}$\downarrow$\\
\midrule
Ground Truth     & 67.63& 0.0& 0.1601& 0.0& 2.819\\
TokenSynth~\citep{kim2025tokensynthtokenbasedneuralsynthesizer}     & 37.99& 31.29& 0.1290& 9.359& 3.055\\
\midrule
No Search  & 23.42& 17.71& 0.0583& 3.977& 1.523\\
Uncond. Search (N=8) & \textbf{18.55}& 16.95& 0.1437& 3.965& 1.346\\
Uncond. Search (N=16) & 22.11& 16.75& 0.1625& 3.976& \textbf{1.336}\\
Uncond. Search (N=32) & 26.06& \textbf{16.65}& \textbf{0.1821}& \textbf{3.832}& 1.385\\
\bottomrule
\end{tabular}%
\end{table}

\subsection{Results}

\textbf{Single-Note Generation.} Table 1 evaluates single-note quality without guided search. 
When evaluating without any sampling by directly taking the predicted mean of the velocity field distribution (the ``No Search'' setting), FlowSynth already surpasses TokenSynth across all metrics but CLAP score, demonstrating that distributional training improves the learned velocity field even when using only the mean prediction. Unconditional search with increasing budgets shows consistent improvements, with N=32 achieving $3.7$\% lower FAD and \textbf{$212$\%} higher CLAP scores compared to deterministic generation. This validates that DFM's learned uncertainty identifies meaningful variation in the generation space.

\textbf{Multi-Note Generation.} 
The power of guided search emerges when optimizing the combined objective:
\begin{equation}
s_{\text{combined}} = \lambda \cdot \text{TCC}(\{x_i\}_{i=1}^M) + (1-\lambda) \cdot (1-\text{CLAP}(x_i, c_{\text{text}})),
\end{equation}
where $\lambda=0.7$ balances consistency with prompt adherence. This guided approach achieves \textbf{25\%} lower TCC than TokenSynth while maintaining competitive CLAP scores, effectively solving the timbre drift problem that limits current virtual instruments.

Table 2 presents results for complete instrument generation across 12 uniformly-spaced pitches (one per octave). The gap between FlowSynth and TokenSynth widens substantially: even without search, FlowSynth achieves \textbf{$85$\%} lower FAD and \textbf{$11$\%} lower TCC, indicating superior authenticity and cross-pitch consistency, respectively. Unconditional search provides modest gains, as random exploration cannot specifically target consistency.

\begin{table}[ht]
\centering

\begin{tabular}{l c c c c}
\toprule
\textbf{Model} & \textbf{F-score}$\uparrow$&  \textbf{CLAP}$\uparrow$& \textbf{FAD}$_{vgg}$$\downarrow$& \textbf{TCC}$\downarrow$\\
\midrule
Ground Truth     & 1.0& 0.1920& 0.0& 1.219\\
TokenSynth     & 0.5999& 0.1560& 10.68& 2.597\\
\midrule
No Search  & \textbf{0.9171}& 0.0942& 1.652& 2.328\\
Uncond. Search (N=8) & 0.9152& 0.1203& \textbf{1.620}& 2.303\\
Uncond. Search (N=16) & 0.9105& 0.1401& 1.677& \textbf{2.297}\\
Uncond. Search (N=32) & 0.9091& \textbf{0.1575}& 1.680& 2.303\\
\bottomrule
\end{tabular}
\caption{Multi-Note Results without guided search}
\end{table}

\subsubsection{Test-Time Scaling Analysis}
For guided single-note generation, we employ CLAP score as the sole search objective:
\begin{equation}
x^* = \arg\max_{x \in \mathcal{X}_N} \text{cos}(\phi_{\text{CLAP}}(x), \phi_{\text{CLAP}}(c_{\text{text}})),
\end{equation}
where $\mathcal{X}_N$ represents N sampled trajectories. This pure prompt-adherence objective improves CLAP scores while maintaining audio quality, confirming that uncertainty-guided exploration effectively navigates the trade-off between fidelity and controllability.

Figure 2 shows how the CLAP score scales with the number of guided steps for a fixed trajectory count ($N=32$). For single-note generation with CLAP-only guidance, performance exhibits a clear \textbf{logarithmic improvement}. Initial steps yield substantial gains in prompt adherence, but the curve quickly plateaus, demonstrating diminishing returns. This confirms that a moderate number of guided steps offers an efficient trade-off, achieving most of the possible quality improvement while minimizing computational cost.

\begin{figure}
    \centering
    \includegraphics[width=0.8\linewidth]{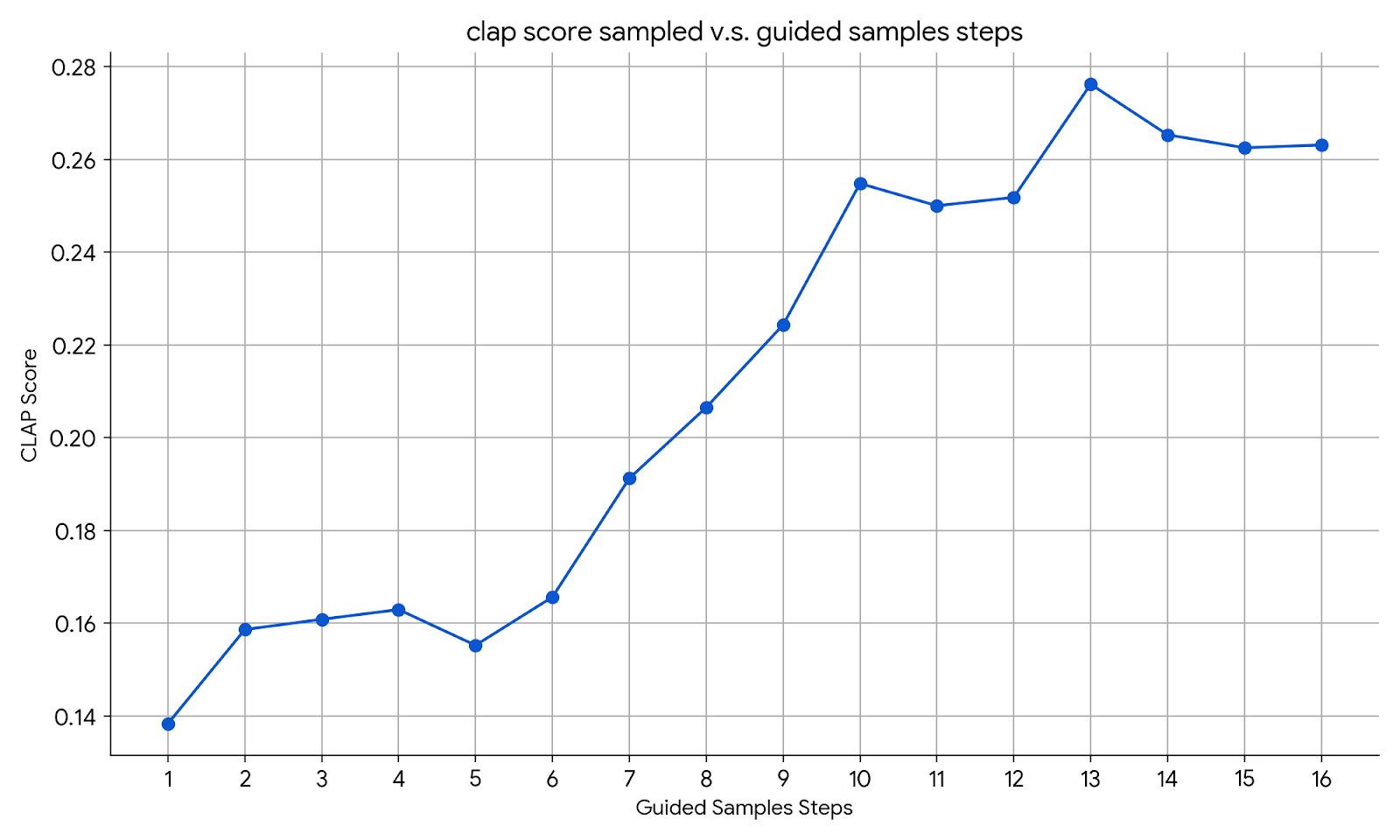}
    \caption{Sample CLAP score v.s. guided sampling steps}
    \label{fig:placeholder}
\end{figure}

The computational overhead scales linearly with N due to parallel evaluation. On a single A100 GPU, generating a complete 88-key instrument takes 45 seconds (deterministic), 1.5 minutes (N=8), or 3 minutes (N=16). This positions FlowSynth as practical for studio use while offering a clear quality-compute trade-off for different production contexts.

\section{Conclusion}
\label{sec:conclusion}

We introduce \textbf{FlowSynth}, a framework that improves virtual instrument timbre consistency using \textbf{Distributional Flow Matching (DFM)} and \textbf{test-time search}. By modeling velocity fields probabilistically, DFM captures predictive uncertainty, enabling an inference-time search that converts computational budget directly into quality gains.

FlowSynth outperforms the state-of-the-art baseline, achieving superior consistency and a natural timbre, as indicated by low Fréchet Audio Distance (FAD) scores. With our guided optimization, performance scales effectively, significantly improving timbre consistency and prompt adherence. This approach demonstrates a practical path toward professional-quality, AI-generated instruments with reliable timbre. Given that we achieved these results on a standard, relatively small dataset, we believe the potential of FlowSynth will be even greater when trained on larger-scale instrument data.

\clearpage
\newpage
\bibliographystyle{plainnat}
\bibliography{paper}

\end{document}